# GRB AFTERGLOW LIGHT CURVES FROM A HYDRODYNAMIC MODEL

E. ZOUAOUI

*University of Sciences and Technology Houari Boumediene (USTHB), Faculty of physics, B.P. 32, El-Alia, 16111 Bab Ezzouar, Algiers, Algeria*
*esma.zouaoui@gmail.com*

M. FOUKA

*Research Centre in Astronomy, Astrophysics and Geophysics, B.P. 63, Algiers observatory, Bouzareah, Algiers, Algeria*
*m.fouka@craag.dz*

S. OUICHAOUI

*University of Sciences and Technology Houari Boumediene (USTHB), Faculty of physics, B.P. 32, El-Alia, 16111 Bab Ezzouar, Algiers, Algeria*
*souichaoui@usthb.dz*

Abstract

In this work, we have modeled the hydrodynamic evolution of the GRBs fireball in violent interaction with the external medium (the ISM) surrounding the burst source by assuming a power law distribution of the accelerated relativistic electrons. For this purpose, a computer code based only on the contribution of the predominant synchrotron radiation mechanism was developed. Light curves for the afterglow emissions following several GRBs over the X-ray and the visible R frequency bands were calculated. Their comparison to observed data by the XRT/Swift satellite and Earth telescopes, respectively, points out fair overall agreements, thus confirming the validity of our hydrodynamic simulation based on the model of Feng et al. (2002).

1. Introduction

Most of the observed gamma ray bursts (GRBs, i.e., very intense flashes of prompt, hard and very brilliant cosmologic electromagnetic radiations) are usually followed by afterglows which consist in remnant, softer, delayed radiations over a large frequency range extending from X-rays down to radio waves. The most popular model describing both two radiation types is the fireball model where two classes of violent collisions are assumed: (i) the internal shocks behind the GRB emission and (ii) the external shock producing the remnant afterglows. In this contribution, we report on a performed hydrodynamic modeling of the external choc that allowed us to calculate afterglows light curves by assuming the predominance of the synchrotron emission mechanism[1], which is justified mainly in the case of low electron densities (typically for $n_e < 10^3$ cm$^{-3}$).

## 2. Hydrodynamic evolution model

For treating the hydrodynamic evolution of the GRB fireball, we have assumed a relativistic jet of conic shape whose wave front remains spherical during all its interaction phases with the external interstellar medium (the ISM) surrounding the burst's source. In this respect, three different models have been studied and compared in details, the models of Chiang and Dermer[2], Huang et al.[3] and Feng et al.[4] The later model, leading to a good solution to a more adequate and realistic hydrodynamic equation, has finally been adopted for describing the afterglows and determining the associated emission energy spectra and light curves. As a first approximation approach, we limited ourselves to the predominant synchrotron radiation as the basic emission mechanism, neglecting other less important absorption and scattering effects such as the self-absorption and inverse Compton radiations, as well as the influence of delayed photons originating from high latitudes. According to the model of Feng and al.[4], the evolution of the fireball is described by the equation

$$\frac{d\Gamma}{dm} = -\frac{\Gamma^2 - 1}{M_0 + m + U_{rel}/c^2 + (1-\varepsilon)\Gamma m}, \quad (1)$$

where $\Gamma$ is the Lorentz factor, c is the velocity of light in vacuum, $M_0$ and m are, respectively, the relativistic jet initial mass and the ISM mass cut off by the fireball, $U_{rel}$ is the energy released by the accelerated electrons within the fireball, and $\varepsilon$ is the fireball radiation efficiency. $U_{rel}$ is given by a general differential formula[4] in term of the internal energy of the shock, $U_{shock}$, i.e.

$$dU_{rel} = (1-\varepsilon)dU_{shock} = (1-\varepsilon)[(\Gamma-1)dmc^2 + mc^2 d\Gamma]. \quad (2)$$

The strength of the model of Feng et al.[4] mainly resides in considering the radiation efficiency as variable due different time scales characterizing the synchrotron radiation and the fireball expansion. Indeed, this quantity can be expressed as[5]

$$\varepsilon = \varepsilon_e \frac{t_{syn}^{-1}}{t_{syn}^{-1} - t_{ex}^{-1}}, \quad (3)$$

where $t_{syn}$ and $t_{ex}$ denote, respectively, the synchrotron cooling time and the expansion time. Indeed, the energy of a radiating electron is lost by two mechanisms, the synchrotron emission and the fireball expansion, which reduces the radiation efficiency of the later.

## 3. Afterglows light curves, results and comment

In the absence of radiation losses, the electron population accelerated by the shock can be described by a power low distribution of the form[1]:

$$N_e(\gamma_e) = \frac{dN_e}{d\gamma_e} = C\gamma_e^{-p}, \quad (4)$$

with the Lorentz energy parameter varying in the range $\gamma_{min} \leq \gamma_e \leq \gamma_{max}$

The synchrotron power for frequency, $\nu'$, is given by

$$P_{\nu'} = \frac{2\sqrt{3}e^2\nu_L}{c} \int_{\gamma_{min}}^{\gamma_{max}} N'_e(\gamma_e) F(x) d\gamma_e, \quad (5)$$

where $\nu_L$ is the Larmor frequency and $F(x)$ is the synchrotron function in the relativistic regime (x being a frequency ratio). Finally, we derived the following general expression for the instantaneous intensity in function of the luminosity distance, $D_L(z)$, which in turn depends[1,6] on the red-shift, z, i.e.:

$$F_\nu = \frac{1}{4\pi D_L(z)^2} 4\pi \frac{dP_\nu}{d\Omega}. \quad (6)$$

Then, integrating over a frequency band [$\nu_1$, $\nu_2$] yields the radiation fluence expressed by

$$S_B = \int_{\nu_1}^{\nu_2} F_\nu d\nu, \quad (7)$$

that can be directly compared to observational data measured by satellite missions or terrestrial telescopes. For the visible and infrared frequency bands, the results are often presented in term of the apparent magnitude defined by

$$M_B = M_\odot - 2.5 \log_{10}\left(\frac{S_B}{W_\odot}\right), \quad (8)$$

where $M_\odot$ = - 26.7 et $W_\odot$ = 1.36×10$^2$ erg s$^{-1}$ cm$^{-2}$ are, respectively, the apparent magnitude and the intensity of solar radiation measured from the Earth.

We assumed a fireball with initial mass outflow, $M_0 = 2\times10^{-6} M_\odot$, initial Lorentz factor $\Gamma_0$ = 250, and a jet opening angle, $\theta_{jet}$= 10°, decelerating within the ISM with constant density, $n$ = 1 cm$^{-3}$, lateral expansion, g = 0, and typical redshift value, z = 1, corresponding to a luminosity distance, $D_L(z)$ = 6.32 Gpc.

In Figs. (1, 2) are plotted light curves calculated using our computer code over the X-ray and the visible R bands, respectively. They are , respectively compared in the same figures to observed light curve data reported by the XRT/Swift satellite instrument over the spectral band E = 0.3-10 keV or measured by terrestrial telescopes in case of the R visible band. As can be noted, general agreements between our hydrodynamic model calculated values and observational data are thus obtained; the apparent differences being likely due to several radiation processes like, e.g., the off-axis emission not included in our calculations.

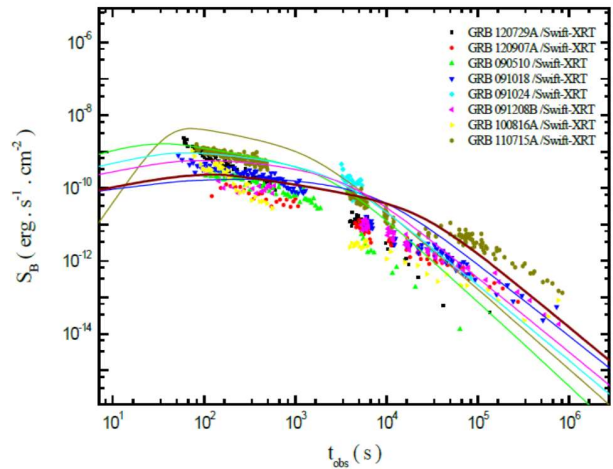

Fig. 1. Comparison of calculated afterglow light curves to observational data reported by the XRT/Swift satellite mission in term of the integrated fluence, $S_B$ (in erg.s$^{-1}$.cm$^{-2}$ units) in the X-ray band (E = 0.2-10 keV).

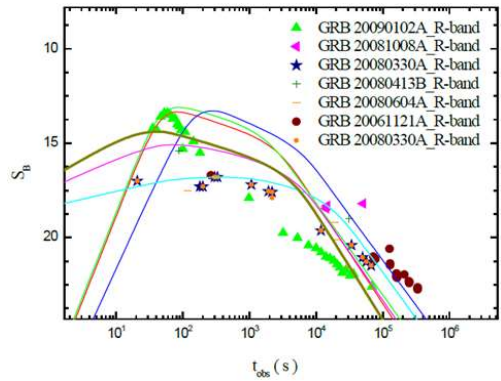

Fig. 2. Comparison of calculated afterglow light curves to observed data in the visible range (the R-band) in term of the apparent magnitude $M_{app}$.